# A Photonics-based superheterodyne RF reception approach

Guangyu Gao, Qijun Liang, Ziyu Liu, Huanfa Peng, Qiang Zhao, Naijin Liu
Qian Xuesen Laboratory of Space Technology,
China Academy of Space Technology, Beijing 100094, China
Email: gaoguangyu@qxslab.cn

**Abstract:** A novel photonics-based RF reception approach is proposed as a competitive solution to meet the current challenges of photonic-based approaches and to realize high performances at the same time. The proposed approach adopts the superheterodyne configuration by a combination manner of electronic techniques and photonic techniques, including the ultra-wideband generation of optical LO, the two-stage photonic superheterodyne frequency conversion and the real-time IF compensation. An engineering prototype has been developed and its performance has been evaluated in the laboratory environment. The experiment results preliminarily verify the feasibility of the proposed approach and its engineering potential. The typical performances are as follows: 0.1 GHz~ 45GHz operation spectrum range (>40 GHz), 900 MHz instantaneous bandwidth, 101 dB·$Hz^{2/3}$ SFDR and 130 dB·Hz LDR, image rejections of ~80 dB for 1st frequency conversion and >90 dB for 2nd frequency conversion.

## 1 Introduction

RF receiver as the front-end in modern electronic system play essential roles in wide range of applications from spectrum sensing, electronic warfare (EW), Radar, to wireless communications. Continuous development is promoting RF receiver to satisfy the increasing requirement on several key performances at the same time, such as wider operation spectrum range (>40 GHz), larger instantaneous bandwidth (>500 MHz), higher interference/spurious suppression capability. Conventional receiver architectures based on the electronic techniques and components have significant technical bottlenecks hardly satisfying these requirements [1-4].

Typically, the superheterodyne configuration is the most common receiver architecture adopted by diverse applications requiring high reception performance, including Radar receiver, EW receiver, test receiver and spectrum/signal analyzer [2, 5-8]. The superheterodyne receiver translates the desired RF signal to one intermediate frequency (IF) through multi-stage conversions and filters before digitization, having excellent reception performance in high imaging rejection, dynamic range, frequency selectivity and sensitivity. Whereas, it encounters four problems: 1) local oscillators (LO) with limited tuning range and low harmonic suppression, 2) RF mixers with limited frequency conversion range and instantaneous bandwidth, 3) tunable preselector with a narrow bandwidth and signal distortion, not compatible with large instantaneous bandwidth, 4) complex structure with multistage conversion leading to poor size, weight, and power consumption (SWaP).

Photonic techniques have attracted great attention in breaking through the bottlenecks of conventional electronic receiver architectures due to its advantages such as ultra-wide operating spectrum range, large instantaneous bandwidth, low nonlinear distortion, low loss, light weight, high-efficiency tunable LO, and high interference/spurious suppression. In recent years, a variety of photonics-based RF reception approaches based on different architectures have been proposed and experimental demonstrated for various application purposes [4, 10-17]. However, three major challenges still need to be addressed. 1) The first challenge is the generation of optical LO. Most of the approaches use electronic LO to generate optical LO by the method of electro-optic modulation (EOM), which brings the problems of electronic LO to the photonic-based RF receiver. For example, the limited frequency tuning range of the electronic LO makes it very difficult to do the frequency plan and parameter designing of the photonic-based the superheterodyne configuration. Furthermore, this method will deteriorate the spurious suppression performance for that EOM of the electronic LO at high driver power will make complex high-order harmonic products. Although some other efforts [16, 18-20] such as optical combs and multiple lasers with injection locking have overcome the challenge to some extent, there exist some new problems in the complexity, reliability, tunability and other aspects. 2) The second challenge is the photonic frequency conversion. Current photonics-based RF reception approaches mainly adopt the direct conversion or low IF configurations [22-25]. On conditions of wide frequency conversion range (typ. 5 GHz~40 GHz) and large instantaneous bandwidth (typ. >500 MHz), these approaches showed better imaging rejections than the pure electronic counterparts, increasing by > 20 dB via the combination method of the photonic IQ down-conversion and the electronic or digital image-rejected processing. But the residue unbalance of the amplitude and phase between I and Q channels makes the imaging rejection performance still lack of competitiveness in comparison with the superheterodyne configuration.

In this paper, we propose and demonstrate a novel photonics-based RF reception approach, which is competitive to meet the current challenges and to realize high performances. The proposed approach adopts the superheterodyne configuration by a combination manner of electronic techniques and photonic techniques, including the ultra-wideband generation of optical LO, the superheterodyne photonic frequency conversion and the real-time IF processing. The photonic-based superheterodyne configuration has two stages of photonic-based frequency down-conversion, where two lasers are respectively used as the optical carrier and the optical LO. The other techniques including digital feed-back locking, analog feed-forward compensation, and RF processing in the frond-end and back-end are employed to optimize the system performance, such as frequency/phase noise compensation, interference/spurs suppression. An engineering prototype has been developed and tested in detail. The proposed approach has two distinguishing features.

1) The optical LO is generated based on a wide-band tuning laser with two major steps, the coarse locking and tuning of the LO frequency by a digital locking unit, and the real-time compensation of the residue frequency-error and the phase-noise by the 2$^{nd}$ frequency conversion unit and the reference extraction unit. The proposed optical LO serve has the advantages of ultra-wide tuning range (>45 GHz), excellent harmonic spurious suppression, and its implementation is relatively simple and reliability.

2) The photonic frequency conversion is based on the superheterodyne configuration. This photonic-based superheterodyne configuration has two stages of frequency conversion. In the first stage, RF signals is converted to a fixed high-intermediate frequency (high-IF). In the second stage, the high-IF signal is down-converted to a low-IF. Combining with the RF processing in the front-end before EOM and the IF processing after the photoelectric detection, the proposed photonic frequency conversion shows several advantages of ultra-wide frequency conversion range (⩾45 GHz), larger instantaneous bandwidth (>900 MHz), higher interference/spurious suppression capability.

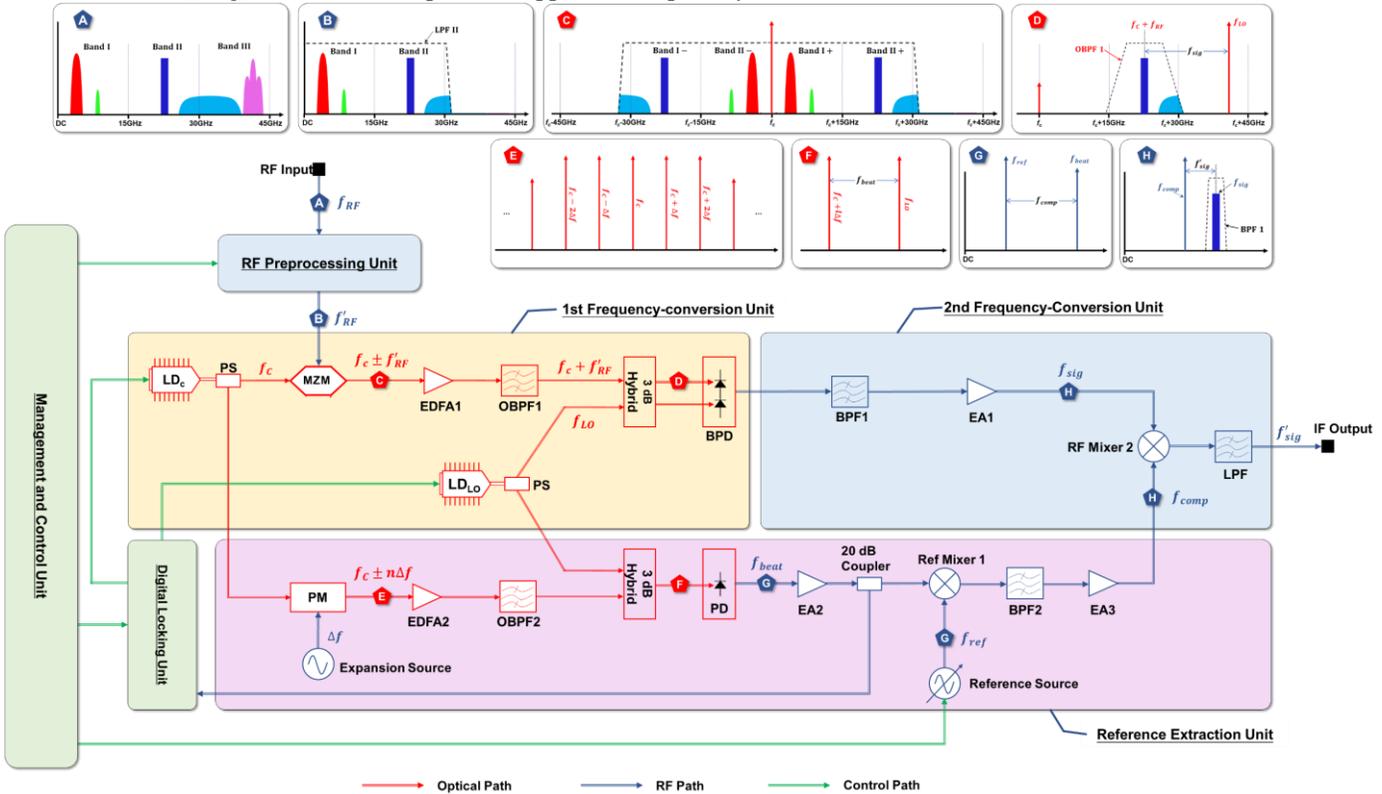

Fig. 1. The schematic of the proposed photonics-based RF reception approach

## 2 Principles

The schematic of the proposed photonics-based RF reception approach is shown in Fig. 1, comprised of six main parts, the RF reprocessing unit, the 1$^{st}$ frequency conversion unit, the 2$^{nd}$ frequency conversion unit, the reference extraction unit, the digital locking unit, and the management and control unit. The first part is the RF reprocessing unit at the RF input-end, the schematic diagram is shown in fig 2. Its main function is to filter the desired signal $f'_{RF}$ (see fig 1B and 2) and to reject the image interference from the input RF signal $f_{RF}$ (see fig 1A), when the system working at a wide operation range (tens of Gigahertz) and a large instantaneous bandwidth (> 500 MHz). For many of the superheterodyne receiver of high performance based on the conventional architectures, such as test receiver and spectrum/signal analyzer, the Yttrium Iron Garnet (YIG) filter are used to preselect desired RF signal due to a wide tuning range up to 40 GHz and a big out-of-band rejection (> 70 dB), while its main disadvantages are a small bandwidth (<50 MHz), a big insertion loss (typ. >5 dB), difficulty in current control and the intrinsic signal distortion. As a result, these receivers should bypass the preselector when working at the mode of wide-band signal receiving and analyzing at the expense of none of the out-of-the-band interferences at the 1st reception stage being suppressed. Thanks to the advantages of the ultra-wideband generation of optical LO and the superheterodyne photonic frequency conversion as mentioned in the introduction, the preselection of the input RF signal can be achieved with only several wideband fixed-frequency filters, which have high selectivity, low cost and small footprint, such as microstrip filter and cavity filter. And for our approach, when working at the wide-band mode, it is not necessary to bypass the preselecting function, for that the filters in this approach are intrinsic wide-band. For example, only two low-pass fixed-frequency microstrip filters, LPF I and LPF II with 3 dB cut-off frequencies at 18 GHz and 32 GHz respectively as shown in fig 2, are used in our approach to achieved an 80 dB image suppression.

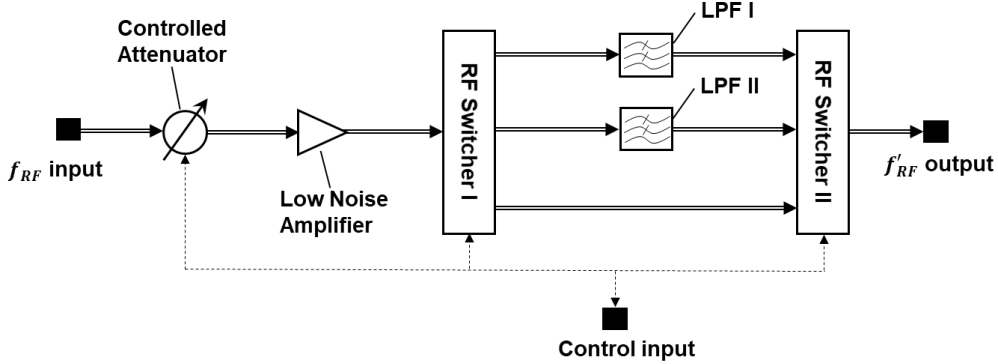

Fig. 2. The schematic of the RF reprocessing unit

In the 1st frequency conversion unit, the desired signal $E'_{RF}(t)$ with frequency at $f'_{RF}$ input from the RF reprocessing unit is up-converted to the optical domain by modulating an optical carrier with frequency at $f_C$ through double-sideband carrier-suppressed modulation (DSB-CS) in a Mach-Zehnder modulator (MZM) as shown in fig 1C. On the assumption that the spectral component at frequency $f'_{RF}$ has the expression $E'_{RF}(t) = V_{f'_{RF}} \sin(2\pi f'_{RF} t)$. Then the $E'_{RF}(t)$ signal modulated on light can be expressed as

$$E_{C\_RF}(t) = e^{j(2\pi f_C t + \varphi_n^C)} \times \sin(\beta \sin(2\pi f'_{RF} t))$$
$$= \sum_{k=1}^{\infty} J_{2k-1}(\beta) \left( e^{j(2\pi(f_C + (2k-1)f'_{RF})t + \varphi_n^C)} - e^{j(2\pi(f_C - (2k-1)f'_{RF})t + \varphi_n^C)} \right) \quad (1)$$

where $\varphi_n^C$ is the phase noise of optical carrier, $\beta$ is the modulation factor given as $\pi V_{f'_{RF}}/V_{\pi\_MZM}$, $J_k(\beta)$ is the $k^{th}$ Bessel function of first kind. $V_{f'_{RF}}$ is the voltage of $f'_{RF}$ and $V_{\pi\_MZM}$ is the half-wave voltage of MZM. At the small signal condition $\beta \ll 1$, $J_{2k-1}(\beta) \sim 0$. Thus, $E_{C-RF}(t)$ is simplified

$$E_{C\_RF}(t) = J_1(\beta) \left( e^{j(2\pi(f_C + f'_{RF})t + \varphi_n^{CC})} - e^{j(2\pi(f_C - f'_{RF})t + \varphi_n^C)} \right) \quad (1\text{-}1)$$

After amplified by the EDFA 1, the upper side band signal with frequency at $f_C + f'_{RF}$ is selected by a fixed-frequency and wideband optical band-pass filter OBPF 1, then

$$E'_{C\_RF}(t) = J_1(\beta) e^{j(2\pi(f_C + f'_{RF})t + \varphi_n^C)} \quad (2)$$

One of the main aims of using OBPF 1 here is to reject the lower-side band of $E_{C-RF}(t)$, the residual optical carrier, the ASE noise from EDFA 1 and the residual image interference on light. For simplification, the OBPF 1 are not used in the experiment validation as descripted in the next section. $E'_{C\_RF}(t)$ and an optical LO with frequency at $f_{LO}$ are coupled in a 3 dB hybrid as shown in fig 1D. The output signals at the two ports of the 3 dB hybrid are

$$\begin{cases} E^1 = J_1(\beta) \left( e^{j(2\pi(f_C + f'_{RF})t + \varphi_n^C)} + e^{j(2\pi(f_{LO} + f_{err})t + \varphi_n^{LO})} \right) \\ E^2 = J_1(\beta) \left( e^{j(2\pi(f_C + f'_{RF})t + \varphi_n^C)} - e^{j(2\pi(f_{LO} + f_{err})t + \varphi_n^{LO})} \right) \end{cases} \quad (3)$$

where $\varphi_n^{LO}$ is the phase noise of optical LO and $f_{err}$ is the frequency error of optical LO relative to optical carrier. The output signals of the 3 dB hybrid are fed into a balance PD for the optical-electronic down-conversion of the $f_C + f'_{RF}$ signal through the coherent beat process. The output current of the balance PD is expressed as

$$E_{sig}(t) \propto E^1 \cdot (E^1)^* - E^2 \cdot (E^2)^*$$
$$\propto 2\cos\left(2\pi\left((f_{LO} + f_{err}) - (f_C + f'_{RF})\right)t + \varphi_n^{LO} - \varphi_n^C\right)$$
$$= 2\cos\left(2\pi\left(f_{sig} + f_{err}\right)t + \Delta\varphi_{n\_C,LO}\right) \quad (4)$$

where $f_{sig} = f_{LO} - (f_C + f'_{RF})$, and $\Delta\varphi_{n\_C,LO} = \varphi_n^{LO} - \varphi_n^C$. In the 1st frequency conversion unit, $f_{sig}$ is set as a constant high-IF, as shown in fig 1H.

For the superheterodyne configuration [6-9], the high-IF frequency should be as high as possible so as to push the image frequency far away from the desired signal for efficient signal selection from the interference by RF filters and as a result increasing the instantaneous bandwidth. Suffering from the bottlenecks of electrical LO in limited tuning range and the RF mixer in limited operation range especially for the LO port, the high-IF frequency in the conventional superheterodyne receiver usually <2 GHz.

These lead to complicating the frequency planning and the system configuration for suppressing the input interferences and the spurious generated in the frequency conversion. Thanks again to the advantages of the ultra-wideband generation of optical LO and the superheterodyne photonic frequency conversion, the high-IF frequency can be chosen at a big value (typ. 18.5 GHz in our approach) so that the optical LO always locates at the upper side of the signal frequency even up to 45 GHz as shown in fig 1D. As a consequence, the 1$^{st}$ frequency conversion with a large instantaneous bandwidth, a high suppression of image frequency can achieve successfully with a considerably simple frequency planning and the system configuration. It should be noted that the optical frequency conversion has an advantage of intrinsic resistance to the spurious in the frequency conversion as compared to the RF mixer [15].

In expression (4), the high-IF signal $E_{sig}(t)$ contains two disturbances, the residue frequency error $f_{err}$ and the residue phase noise $\Delta\varphi_{n\_C,LO}$. Although the optical LO laser is coarse frequency-locked (not phase-locked) to the optical carrier laser in a digital feed-back locking manner (details described in the following part), the 1$^{st}$ photonic frequency conversion in our approach will introduces obvious frequency instability and phase noise to the high-IF signal $f_{sig}$ due to the frequency instability and the noncoherence between the two lasers. In order to address these issues, the analog feed-forward compensation method is introduced. The compensation signal is generated in the reference extraction unit, the schematic of which is shown in fig 1 (see the pink block). In this unit, one copy of the optical carrier is modulated with a phase modulator (PM) driven by an electrical expansion source with a fixed-frequency at $\Delta f$. The output signal after the phase modulation is

$$E_{C\_\Delta f}(t) = e^{j(2\pi f_C t + \varphi_n^C + \alpha \sin(2\pi \Delta f t + \varphi_n^E))}$$
$$= \sum_{-\infty}^{+\infty} J_l(\alpha) e^{j\left(2\pi(f_C + l\Delta f)t + \varphi_n^C + l\varphi_n^E + l\frac{\pi}{2}\right)} \quad (5)$$

where $\varphi_n^E$ is the phase noise of the expansion source, $\alpha$ is the modulation factor given as $\pi V_{\Delta f}/V_{\pi\_PM}$, $J_l(\beta)$ is the $l^{th}$ bessel function of first kind. $V_{\Delta f}$ is the voltage of $\Delta f$ and $V_{\pi\_PM}$ is the half-wave voltage of PM. By properly setting $V_{\Delta f}$, several harmonic modes of $f_C$ with frequencies at $f_C \pm l\Delta f$ as shown in fig 1E are generated with relatively flat power distribution when $|l| \leq 2$. After amplified by the EDFA 2 and filtering by a fixed-frequency optical band-pass filter OBPF 2, only one harmonic mode with the mode number $2 \geq l \geq 0$ is selected and coupled with one copy of the optical LO in a 3 dB hybrid as shown in fig 1F. The output signal of the 3 dB hybrid is

$$E^3 = e^{j(2\pi(f_{LO}+f_{err})t+\varphi_n^{LO})} + J_l(\alpha) e^{j\left(2\pi(f_C+l\Delta f)t+\varphi_n^C+l\varphi_n^E+l\frac{\pi}{2}\right)} \quad (6)$$

It is fed into a single PD to create a beating signal with frequency at $f_{beat}$. The beating signal is used for the compensation signal extraction. The output current of the single PD (SPD) is

$$E_{beat}(t) \propto E^3 \cdot (E^3)^*$$
$$\propto 2\cos\left(2\pi((f_{LO}+f_{err})-(f_C+l\Delta f))t + \varphi_n^{LO} - \varphi_n^C - l\varphi_n^E - l\frac{\pi}{2}\right)$$
$$= 2\cos\left(2\pi(f_{beat}+f_{err})t + \Delta\varphi_{n\_C,LO} - l\varphi_n^E - l\frac{\pi}{2}\right) \quad (7)$$

where $f_{beat} = f_{LO} - (f_C + l\Delta f)$. Apparently, $E_{beat}(t)$ contains the same information of the two disturbances in $E_{sig}(t)$, $f_{err}$ and $\Delta\varphi_{n\_C,LO}$. In the reference extraction unit, $f_{beat}$ is set to satisfy the following relationship

$$f_{beat} \in \left(\min(f_{LO}) - f_C, \max(f_{LO}) - f_C - \max(l) \times \Delta f\right] \quad (8)$$

where $\min(*)$ and $\max(*)$ are the minimum and the maximum values of *. By mixing with a reference signal $E_{ref}(t)$ from a tunable reference source with frequency at $f_{ref}$, the beating signal $f_{beat}$ is converted to a constant IF signal $E_{comp}(t)$ with frequency at $f_{comp}$ as shown in fig 1G, which serving as the compensation signal for reducing the residue frequency error and the residue phase noise of $E_{sig}(t)$. $f_{ref}$ is set to satisfy the following relationship

$$f_{ref} \in \left(\min(f_{LO}) - f_C, \max(f_{LO}) - f_C - \max(l) \times \Delta f\right] \quad (9)$$

$E_{comp}(t)$ can be given as

$$E_{comp}(t) = E_{comp} = \frac{1}{2}\cos\left(2\pi\left(f_{comp}+f_{err}\right)t + \Delta\varphi_{n\_C,LO} - l\varphi_n^E - \varphi_n^{ref} - l\frac{\pi}{2}\right) \quad (10)$$

where $f_{comp} = f_{beat} - f_{ref}$. $E_{comp}(t)$ is transferred to the 2$^{nd}$ frequency conversion unit, wherein $E_{comp}(t)$ also serves as the 2$^{nd}$ LO to down-convert $E_{sig}(t)$ to a low-IF signal $E'_{sig}(t)$ with frequency at $f'_{sig} = f_{sig} - f_{comp}$ in the RF mixer 2. $E'_{sig}(t)$ as shown in fig 1H can be expressed as

$$E'_{sig}(t) = LPF\left(E_{sig}(t) \times E_{comp}\right) = \frac{1}{2}\cos\left(2\pi\left(f_{sig}-f_{comp}\right)t + l\varphi_n^E + \varphi_n^{ref} + l\frac{\pi}{2}\right)$$
$$= \cos\left(2\pi f'_{sig} t + l\varphi_n^E + \varphi_n^{ref} + l\frac{\pi}{2}\right)$$
$$= \cos\left(2\pi\left(f_{ref}+l\Delta f - f'_{RF}\right)t + l\varphi_n^E + \varphi_n^{ref} + l\frac{\pi}{2}\right) \quad (11)$$

According to (11), the output signal $E'_{sig}(t)$ contains the information of the input signal $f'_{RF}$, and the residual frequency error $f_{err}$ and the residual phase noise $\Delta\varphi_{n_C,LO}$ are all eliminated. While, there are two additional phase noise terms in (11), $l\varphi_n^E$ and $\varphi_n^{ref}$, which are introduced in the process of compensation. As for $l\varphi_n^E$, it is originated from a single tone with a fixed-frequency at $\Delta f$, but has neglectable influence on $E'_{sig}(t)$. The main reason is that the single tone with frequency on 10 GHz level (typ. 15 GHz in this work) and ultra-low phase noise and high frequency stability is easy to achieve and has a relatively low-cost and a small footprint, furthermore the mode number $l$ is so small that has at most 6 dB influence on $E'_{sig}(t)$. As for $\varphi_n^{ref}$, it is originated from the tunable reference source with frequency at $f_{ref}$ and has indistinct influence on $E'_{sig}(t)$. Due to $f_{ref}$ only depending on the value of $f_{beat}$ as descripted in (8) and (9), $f_{ref}$ has a limited tuning range typically covering 15 GHz and $f_{ref} \in [1.5 GHz, 16.5 GHz]$. This type of tunable source with a phase noise lower than -115 dBc/Hz (@ 10kHz offset and 10 GHz central frequency) is available and has a relatively low-cost and a small footprint.

The schematic of digital locking unit is shown in fig. 3, which is based on the offset-frequency discrimination and the digital feed-back decision control loop. In order to generate an RF beat-note with frequency within the bandwidth of the PD, the relative frequency difference $f_{beat}$ between the optical LO laser and optical carrier laser is initialed to be around 18.5 GHz (corresponding to the zero point of $f_{sig}$ in the 1st frequency conversion unit) via thermal tuning the wavelength of the lasers. The $f_{beat}$ signal is sent to a RF frequency divider (FD). Then a low pass filter is used to abstract the 1st order of the frequency-divided signal with a frequency of $f_{beat}/N$. A double-balanced mixer (DBM) is used as a frequency discriminator, which generates an error signal through converting the output of the RF frequency divider by a frequency reference to a fixed frequency containing a frequency error information. After filtering out the high frequency components including the sum frequency and the other spurs, the error signal is sampled by an ADC and the error frequency is calculated by a FPGA-based program. According to the error frequency, the control signal is fed back to drive the TEC and PZT of the lasers, then the relative frequency difference $f_{beat}$ is stabilized by tracking the frequency of the LO laser to the frequency of the carrier laser. Once $f_{beat}$ is coarsely locked to 18.5 GHz, the agility mode is switched on, by which $f_{beat}$ is rapidly tuned by changing the control voltage on the PZT of the lasers, corresponding to the reception frequency of $f_{RF} = f_{beat} - 18.5$ GHz$+ l\Delta f$.

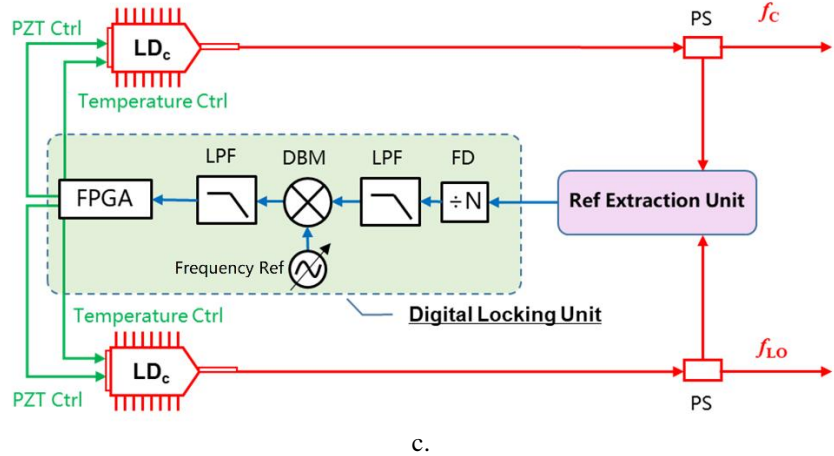

c.

Figure 3. Schematic of the digital locking unit. FD: Frequency divider, LPF: low pass filter, DBM: double-balanced mixer, RF ref: RF reference.

The main function of the management and control unit is to monitor and control each functional part as shown in fig 1. For example, it will transfer the frequency parameter of the desired input signal from the user interface to the digital locking unit for tuning the frequency of optical LO. A customized industrial computer module is used as the management and control unit, in which the user interface and the background program are developed based on the LabVIEW environment.

## 3  Results and discussion

An engineering prototype based on the proposed approach has been developed as shown in fig. 4. The performance evaluation of the engineering prototype is carried out in detail in the laboratory environment. The experiment results as reported in the following parts preliminarily verify the feasibility of the proposed approach and its engineering potential.

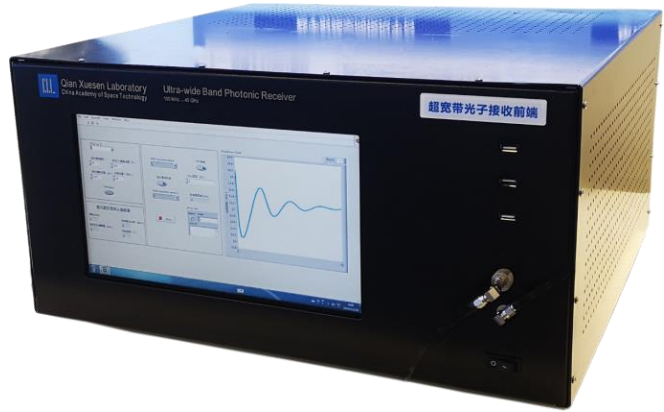

Figure 4 the engineering prototype of the proposed photonics-based RF reception approach

At the input end of the photonics-based RF receiver, the RF spectrum from DC to 45 GHz is divided into 3 bands as shown in fig 1A, and the band I and band II are respectively selected in the RF reprocessing unit by two low-pass fixed-frequency microstrip filters with 3 dB cut-off frequencies at 18 GHz and 32 GHz, LPF I (Marki FLP-1800) and LPF II (Marki FLP-3200). Due to the value of the high-IF $f_{sig}$ is set at 18.5 GHz and the optical LO locates at the upper-side of the modulation signal on light ($f_C + f'_{RF}$), the image interferences of band I and band II with frequency at $f'_{RF} + 37$ GHz are all significantly rejected by LPF I and LPF II respectively with an out-band rejection ratio at ~80 dB. Two narrow-linewidth fiber lasers (NKT basik E15) with the center wavelengths at 1550.12 nm and a linewidth at 100 Hz and tuning range 100 GHz (temperature tuning) and 8 G (PZT tuning) are used to generate optical carrier and optical LO. The MZM (iXblue MXAN-LN-40) with a 3 dB bandwidth > 28 GHz and $V_{\pi\_MZM}$ < 6 V has an excellent suppression ratio of even-order components at > 30 dB for the optical carrier and ~65 dBc for the second-order harmonics when the input power of $f'_{RF}$ at 0 dBm, far better than the wide-band RF mixer by about 20 dB. The BPD (Finisar BPDV2150R) and SPD (Finisar HPDV2120R) with the bandwidths >40 GHz and 50 GHz are used for optical-electronic down-conversion. A tunable optical band-pass filter (Teraxion TFN) with bandwidth at 10 GHz and working at the fixed-frequency mode is used as OBPF 2 for optical filtering in the reference extraction unit, which has a roll-off factor of >600 GHz/nm and out-of-band rejection of around 40 dB. An electrical spectrum analyzer (R&S FSW43) and a real-time oscilloscope (Keysight DSA91304A) are used to measure the electrical spectra and record the signal waveform.

The generation of optical LO is implemented by the procedure as following, and is evaluated by the performances of the frequency drift as well as the other supporting parameters, such as the frequency stability and the phase noise. The frequency stability of the narrow-linewidth fiber laser in free running has frequency drifts on the order of 1 GHz per day and 1 MHz per second, dramatically affecting the frequency accuracy and precision of the photonic receiver. Due to phase-locked loop (PLL) having a small locking bandwidth and a weak robustness for the phase locking of the optical LO laser to the optical carrier when working on an ultra-wideband tuning mode, a coarse frequency-locked manner based on the digital feed-back locking is introduced to achieve a primary frequency control of high-robustness. The coarse frequency locking stabilizes the relative frequency difference $f_{beat}$ between optical LO and optical carrier and is beneficial to reduce the design difficulties in frequency plan and component parameters, such as Low IF and the optical/electrical filters, especially in the processes of the 2nd frequency conversion and the compensation signal extraction. The input frequency range of the frequency divider is 500 MHz~40 GHz with division ratio of 125, so as to cover the full rapidly tunable frequency range (18.5 GHz~33.5 GHz) of $f_{beat}$ and provide enough margin for initialization. The signal after the frequency divider passes through a low pass filter with 3dB cut-off frequency of around 280 MHz, so as to abstract the first order signal and suppress high order signals. The output of the RF frequency divider and a microwave reference is mixed and discriminated in the DBM. The output error signal of the DBM passes through a second low pass filter with 3dB cut-off frequency of 100 MHz, and then is sent to FPGA for real-time frequency estimation and digital feed-back controlling. Ensuring that the beat-note frequency can be effectively controlled at both positive and negative directions, we chose 50 MHz as the determining condition of the frequency of the error signal. Once the frequency of the RF reference is set to be $f_{ref}$, the frequency difference between the LO laser and reference carrier laser is rapidly locked at ($f_{ref}$ +50 MHz)×125. After the coarse frequency locking, the long-term frequency drift of the relative frequency difference $f_{beat}$ is reduced to a small range of about $f_{err} \leq 5$ MHz. The instantaneous frequency $f_{beat}$ of the beating signal $E_{beat}(t)$ is measured using the electrical spectrum analyzer, and the results of the lasers in free running and digital locked are depicted in fig 5. In fig. 5A, the frequency drift of $f_{beat}$ without digital locking is more than 100 MHz on 5 minutes. After digital locking as shown in fig. 5B, the frequency drift decreases to less than 10 MHz in a long term over several hours. This value mainly depends on the frequency threshold parameter set in the management and control unit and frequency controlling precision in the digital locking unit.

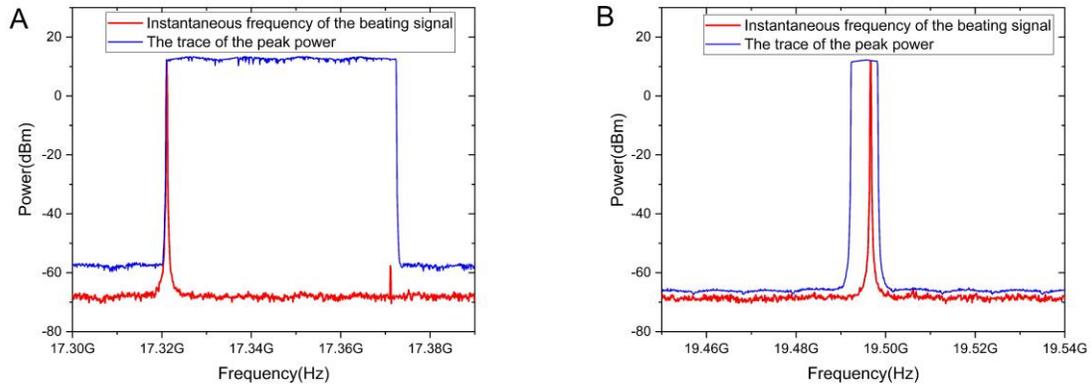

Figure 5 the frequency drift of the beating signal between optical LO and optical carrier before (A) and after (B) frequency locked

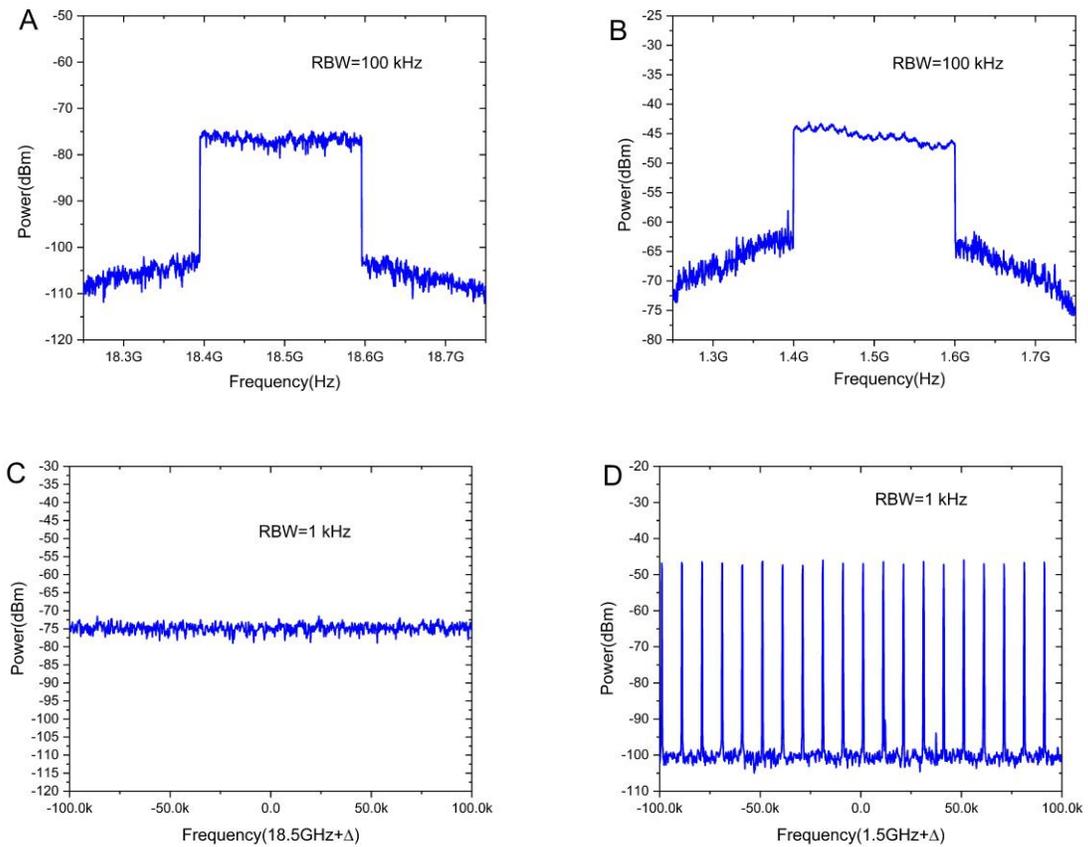

Figure 6 the spectral purity of the RF input signals before (A, C) and after (B, D) the analog feed-forward compensation. C and D are the zoom-in view of A and B, respectively.

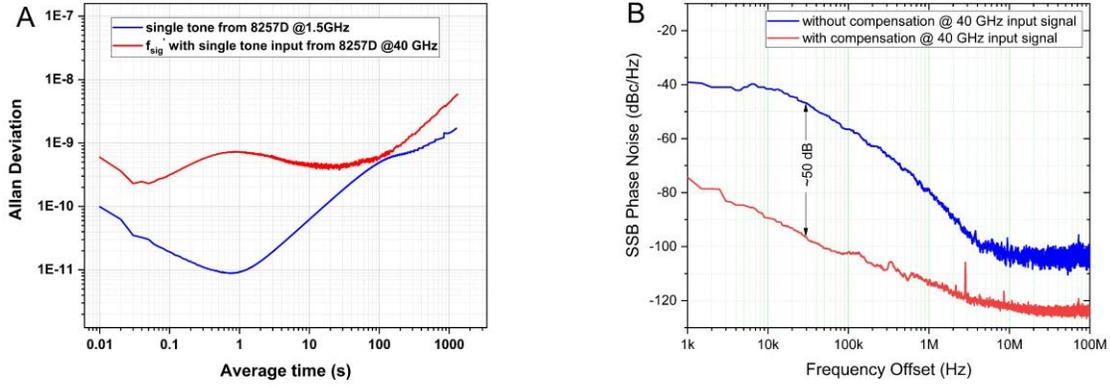

Figure 7. the Allan variance (A) and the phase noise (B) performances of the proposed approach

In order to quantify the effectiveness of the proposed analog feed-forward compensation method, we have analyzed the spectral purity of the RF input signals after down-conversion to IF, before and after the analog feed-forward compensation. The results are shown in Fig. 6. Fig. 6A depicts the power spectrum of a multi-tone signal with spacing at 10 kHz generated from an arbitrary waveform generator (Keysight M8190A), which is measured directly after the BPD by the frequency spectrum analyzer, where its corresponding signal $f_{sig}$ is not compensated. Fig. 6C is the zoom-in view of the multi-tone signal, where the comb-lines are indistinguishable. Fig. 6B depicts the power spectrum of the multi-tone signal after the compensation in the 2$^{nd}$ frequency conversion unit. From its zoom-in view shown in fig. 6D, the comb-lines are distinguishable. According to the results shown in fig. 6, it is obvious that the compensation significantly optimizes the quality of signal output from the 1$^{st}$ frequency conversion unit. The spectral purity is further evaluated by calculating the Allan variance and the phase noise as shown in fig. 7. For calculating the Allan variance, a single tone signal with frequency at 40 GHz generated by a RF source (Keysight 8257D) is down-converted to the low-IF $f'_{sig}$ at 1.5GHz. The corresponding $f'_{sig}$ frequency is measured by a frequency counter (Keysight 53220A). The red line in fig. 7A is the calculated Allan variance of the low-IF $f'_{sig}$ versus the averaging time. The Allan variance is gradually going high from 0.1s to 1s, and slowly decreasing from 1s to 30s, and then rapidly increasing with averaging time after 30s. At one second average, the Allan variance is less than 10$^{-9}$, which is determined by both the extension RF source in the reference extraction unit and the compensation RF source in the 2$^{nd}$ frequency conversion unit. Due to the poor frequency stability before the compensation, the Allan variance of the output signal from the 1$^{st}$ frequency conversion unit is not measured. The result of the Allan variance indicates that the analog feed-forward compensation significantly improves the frequency stability by a factor around 1e5 Hz for 1s averaging time. As a comparison, the Allan variance of one single tone signal with frequency at 1.5 GHz from the same RF source (Keysight 8257D) is shown in fig. 7A (the blue line). The single-side band phase noise (SSB phase noise) of the same single tone signal with frequency at 40 GHz before and after the analog feed-forward compensation are measured as shown in fig. 7B. The waveforms of the high-IF signal $f_{sig}$ at around 18.5 GHz and the low-IF signal $f'_{sig}$ at 1.5GHz are respectively recorded by the real-time oscilloscope and then their SSB phase noises are calculated through a digital signal processing algorithm. As shown in fig. 7B, the SSB phase noise before the compensation (blue line) is above -60 dBc/Hz at the frequency offset below 100 kHz. The high phase noise at the low frequency offset reveals a bad frequency stability which makes the main contribution to the difficulty in phase locking. After the feed-forward compensation (see red line in fig. 7B), the SSB phase noise is significantly reduced with the maximum at around 50 dB, to ~-103 dBc/Hz at the frequency offset 100 kHz. This result confirms the effectiveness of the analog feed-forward compensation.

The system performances are also characterized in terms of dynamic range and linearity as shown in fig. 8. Two tones with frequencies at $f_1$=39.9 GHz and $f_2$= 40 GHz generated respectively from Keysight 8257D and Keysight 8267D are used to measure the fundamental and the third-order intermodulation (IM3). In fig. 8A, the blue line and the red line depict respectively the linear response of the fundamental signal $f_1$ and $f_2$, and the third-order response of $2 \times f_1 - f_2$ and $2 \times f_2 - f_1$, and the spurious-free dynamic range (SFDR) is calculated near to 101 dB·Hz$^{2/3}$ with the noise floor at ~-142 dBm/Hz. The distribution ranges of SFDR and 1dB compression point (P1dB) over the 40 GHz operation range are equal to 2.3 dB and 3.4 dB, as shown in fig. 8B and 8C, indicating a good uniformity of the proposed approach over an ultra-wideband operation range. The linear dynamic range (LDR) at 40 GHz is estimated to be around 130 dB·Hz. The performances of SFDR, P1dB and the LDR of this approach are mainly limited by the dynamic range of the MZM modulator and the EDFA, and the noise contributions from ASE noise and BPD.

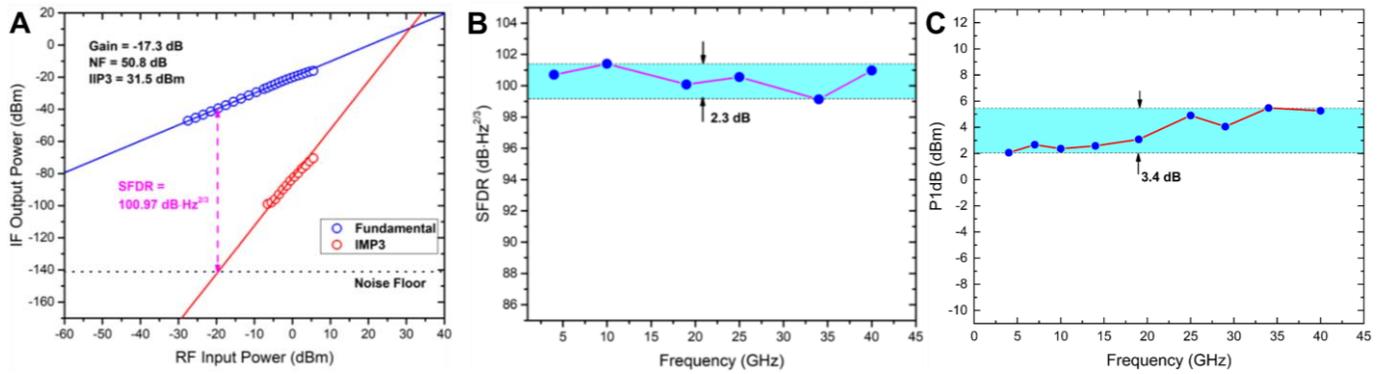

Fig. 8 A: the measured fundamental and IMD3 versus the input RF power. B and C: the SFDR and P1dB distributions over the whole operation range.

In order to evaluate the wideband reception performance of the proposed approach, a 16QAM wideband vector signal generated from the microwave signal generator with carrier frequency at 4 GHz and a symbol rate of 100 MSymbol/s and power of -5 dBm is down-converted and demodulated for the performance evaluation of the back-to-back reception. The low-IF power spectrum of the wideband signal and the constellation diagram are shown in Fig. 9A and 9B. A clear constellation diagram is observed, and the calculated error vector magnitude (EVM) is <1.7%, indicating a good wideband reception performance of the proposed photonic receiver. In fact, the instantaneous bandwidth of this approach mainly depends on the high-IF filter BPF1, which has a 3 dB flat-top bandwidth of 900 MHz and a fixed center frequency of 18.5 GHz. If a bandwidth tunable filter is employed as BPF1, then the instantaneous bandwidth could be reconfigurable.

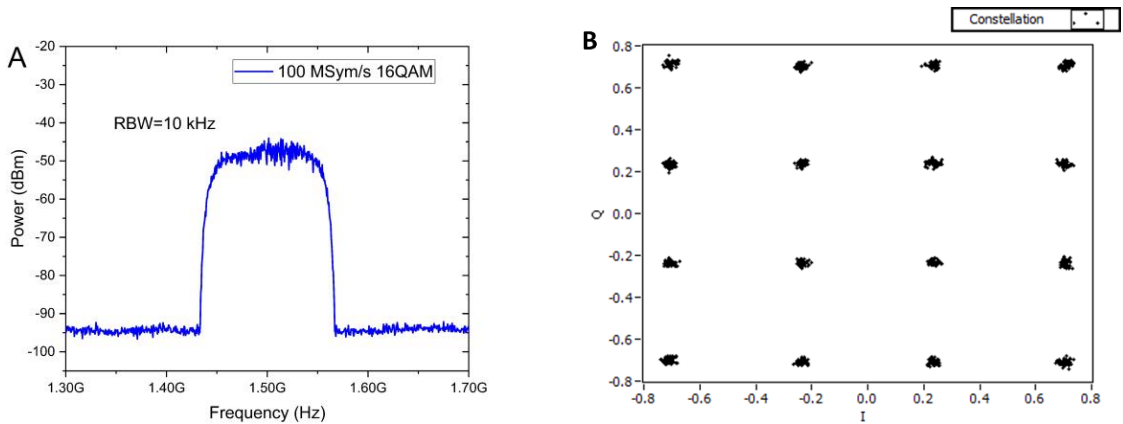

Fig.9. Constellation diagrams demodulated by subharmonic I/Q modulator

Two types of the image interferences exist in this approach, and they are both out of the instantaneous bandwidth as differring from the direct conversion or low IF configurations. One image interferences is the far-end interference locating at the frequency 2×18.5 GHz away from the desired signal, and the other is the near-end interference locating at the frequency 2×1.5 GHz away from the desired signal. Thus, the rejection performances for the two types of image interferences are evaluated. The result shown in fig. 10A is the image rejection of the 1st frequency conversion. From 1 GHz to 7 GHz, the image rejections are around 80 dB. Due to the constraint of RF source (Keysight 8267D) which has a maximum output frequency at 44 GHz, the image rejection in the operation band above 7 GHz is not measured. In practicality, due to the frequency response constraints of RF components at image frequency above 45 GHz, such as antenna, LNA, switcher, to measure the image rejection above 7 GHz make little sense for the 1st frequency conversion. Fig. 10B is the image rejection of the 2nd frequency conversion. At several typical frequencies covering the whole operation range, the image rejections are > 90 dB. It should be noted that these measured values of the two types of image rejections are on the condition that only two RF filters are respectively used in the RF reprocessing unit as a wideband low-pass preselector and in the 2nd frequency conversion unit as a wideband band-pass IF filter, and no optical bandpass filter is used in the 1st frequency conversion unit in the experiment. The image rejections can be further improved to a large extent by connecting an extra RF filter with the same parameters in series or an optical bandpass filter after the EDFA 1 in the 1st frequency conversion unit.

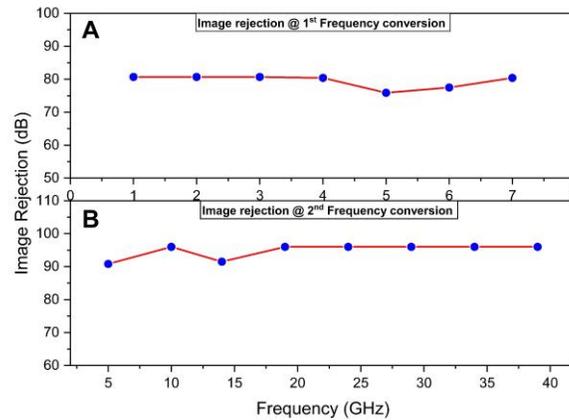

Fig. 10 Measured image rejections for different frequencies of the input RF signal.

## 4 Conclusion

In conclusion, a novel photonics-based RF reception approach is proposed as a competitive solution to meet the current challenges of photonic-based approaches and to realize high performances at the same time. The proposed approach adopts the superheterodyne configuration by a combination manner of electronic techniques and photonic techniques, including the ultra-wideband generation of optical LO, the superheterodyne photonic frequency conversion and the real-time IF compensation. The principle and the functional parts of the proposed approach are described in detail. An engineering prototype of the photonics-based RF reception approach has been developed and its performance has been evaluated in the laboratory environment. The experiment results preliminarily verify the feasibility of the proposed approach and its engineering potential. The typical performances are as follows: 0.1 GHz~ 45GHz operation spectrum range (>40 GHz), 900 MHz instantaneous bandwidth, 101 dB·Hz$^{2/3}$ SFDR and 130 dB·Hz LDR, image rejections of ~80 dB for 1$^{st}$ frequency conversion and >90 dB for 2$^{nd}$ frequency conversion. The proposed approach has two distinguishing features.

1) The optical LO is generated based on a wide-band tuning laser with two major steps, the coarse locking and tuning of the LO frequency by a digital locking unit, and the real-time compensation of the residue frequency-error and the phase-noise by the 2$^{nd}$ frequency conversion unit and the reference extraction unit. The proposed optical LO serve has the advantages of ultra-wide tuning range (>45 GHz), excellent harmonic spurious suppression, and its implementation is relatively simple and reliability.
2) The photonic frequency conversion is based on the superheterodyne configuration. This photonic-based superheterodyne configuration has two stages of frequency conversion. In the first stage, RF signals is converted to a fixed high-intermediate frequency. In the second stage, the high-IF signal is down-converted to a low-IF. Combining with the RF processing in the front-end before EOM and the IF processing after the photoelectric detection, the proposed photonic frequency conversion shows several advantages of ultra-wide frequency conversion range (≥45 GHz), larger instantaneous bandwidth (>900 MHz), higher interference/spurious suppression capability.

## Acknowledgment

This work was partially supported by Independent Innovation Fund of Qian Xuesen Laboratory of Space Technology, and Independent research and development projects of China Aerospace Science and Technology Corporation.